\documentstyle[twoside]{article}
\input{epsf} 
\newcommand\jhep[3]{JHEP {\bf #1}, #3 (#2)} 
\newcommand\npb[3]{Nucl.\ Phys.\ B {\bf #1}, #3 (#2)} 
\newcommand\plb[3]{Phys.\ Lett.\ B {\bf #1}, #3 (#2)} 
\newcommand\Prd[3]{Phys.\ Rev.\ D {\bf #1}, #3 (#2)}
\newcommand\Prl[3]{Phys.\ Rev.\ Lett.\ {\bf #1}, #3 (#2)}

\catcode`\@=11
\long\def\@makefntext#1{
\protect\noindent \hbox to 3.2pt {\hskip-.9pt  
$^{{\eightrm\@thefnmark}}$\hfil}#1\hfill}		

\def\@makefnmark{\hbox to 0pt{$^{\@thefnmark}$\hss}}	
	
\def\ps@myheadings{\let\@mkboth\@gobbletwo
\def\@oddhead{\hbox{}
\rightmark\hfil\eightrm\thepage}   
\def\@oddfoot{}\def\@evenhead{\eightrm\thepage\hfil
\leftmark\hbox{}}\def\@evenfoot{}
\def\sectionmark##1{}\def\subsectionmark##1{}}

\oddsidemargin=\evensidemargin
\newcounter{sectionc}\newcounter{subsectionc}\newcounter{subsubsectionc}
\renewcommand{\section}[1] {\vspace{12pt}\addtocounter{sectionc}{1} 
\setcounter{subsectionc}{0}\setcounter{subsubsectionc}{0}\noindent 
	{\tenbf\thesectionc. #1}\par\vspace{5pt}}
\renewcommand{\subsection}[1] {\vspace{12pt}\addtocounter{subsectionc}{1} 
	\setcounter{subsubsectionc}{0}\noindent 
	{\bf\thesectionc.\thesubsectionc. {\kern1pt \bfit #1}}\par\vspace{5pt}}
\renewcommand{\subsubsection}[1] {\vspace{12pt}\addtocounter{subsubsectionc}{1}
	\noindent{\tenrm\thesectionc.\thesubsectionc.\thesubsubsectionc.
	{\kern1pt \tenit #1}}\par\vspace{5pt}}
\newcommand{\nonumsection}[1] {\vspace{12pt}\noindent{\tenbf #1}
	\par\vspace{5pt}}

\newcounter{appendixc}
\newcounter{subappendixc}[appendixc]
\newcounter{subsubappendixc}[subappendixc]
\renewcommand{\thesubappendixc}{\Alph{appendixc}.\arabic{subappendixc}}
\renewcommand{\thesubsubappendixc}
	{\Alph{appendixc}.\arabic{subappendixc}.\arabic{subsubappendixc}}

\renewcommand{\appendix}[1] {\vspace{12pt}
        \refstepcounter{appendixc}
        \setcounter{figure}{0}
        \setcounter{table}{0}
        \setcounter{lemma}{0}
        \setcounter{theorem}{0}
        \setcounter{corollary}{0}
        \setcounter{definition}{0}
        \setcounter{equation}{0}
        \renewcommand{\thefigure}{\Alph{appendixc}.\arabic{figure}}
        \renewcommand{\thetable}{\Alph{appendixc}.\arabic{table}}
        \renewcommand{\theappendixc}{\Alph{appendixc}}
        \renewcommand{\thelemma}{\Alph{appendixc}.\arabic{lemma}}
        \renewcommand{\thetheorem}{\Alph{appendixc}.\arabic{theorem}}
        \renewcommand{\thedefinition}{\Alph{appendixc}.\arabic{definition}}
        \renewcommand{\thecorollary}{\Alph{appendixc}.\arabic{corollary}}
        \renewcommand{\theequation}{\Alph{appendixc}.\arabic{equation}}
        \noindent{\tenbf Appendix \theappendixc #1}\par\vspace{5pt}}
\newcommand{\subappendix}[1] {\vspace{12pt}
        \refstepcounter{subappendixc}
        \noindent{\bf Appendix \thesubappendixc. {\kern1pt \bfit #1}}
	\par\vspace{5pt}}
\newcommand{\subsubappendix}[1] {\vspace{12pt}
        \refstepcounter{subsubappendixc}
        \noindent{\rm Appendix \thesubsubappendixc. {\kern1pt \tenit #1}}
	\par\vspace{5pt}}

\newcommand{\textlineskip}{\baselineskip=13pt}
\newcommand{\smalllineskip}{\baselineskip=10pt}

\def\eightcirc{
\begin{picture}(0,0)
\put(4.4,1.8){\circle{6.5}}
\end{picture}}
\def\eightcopyright{\eightcirc\kern2.7pt\hbox{\eightrm c}} 

\newcommand{\copyrightheading}[1]
	{\vspace*{-2.5cm}\smalllineskip{\flushleft
	{\footnotesize International Journal of Modern Physics A, #1}\\
	{\footnotesize $\eightcopyright$\, World Scientific Publishing
	 Company}\\
	 }}

\def\abstracts#1#2#3{{
	\centering{\begin{minipage}{4.5in}\baselineskip=10pt\footnotesize
	\parindent=0pt #1\par 
	\parindent=15pt #2\par
	\parindent=15pt #3
	\end{minipage}}\par}} 

\renewenvironment{thebibliography}[1]
	{\frenchspacing
	 \ninerm\baselineskip=11pt
	 \begin{list}{\arabic{enumi}.}
	{\usecounter{enumi}\setlength{\parsep}{0pt}
	 \setlength{\leftmargin 12.7pt}{\rightmargin 0pt} 
	 \setlength{\itemsep}{0pt} \settowidth
	{\labelwidth}{#1.}\sloppy}}{\end{list}}

\newcounter{itemlistc}
\newcounter{romanlistc}
\newcounter{alphlistc}
\newcounter{arabiclistc}

\newcommand{\fcaption}[1]{
        \refstepcounter{figure}
        \setbox\@tempboxa = \hbox{\footnotesize Fig.~\thefigure. #1}
        \ifdim \wd\@tempboxa > 5in
           {\begin{center}
        \parbox{5in}{\footnotesize\smalllineskip Fig.~\thefigure. #1}
            \end{center}}
        \else
             {\begin{center}
             {\footnotesize Fig.~\thefigure. #1}
              \end{center}}
        \fi}

\newcommand{\tcaption}[1]{
        \refstepcounter{table}
        \setbox\@tempboxa = \hbox{\footnotesize Table~\thetable. #1}
        \ifdim \wd\@tempboxa > 5in
           {\begin{center}
        \parbox{5in}{\footnotesize\smalllineskip Table~\thetable. #1}
            \end{center}}
        \else
             {\begin{center}
             {\footnotesize Table~\thetable. #1}
              \end{center}}
        \fi}

\def\@citex[#1]#2{\if@filesw\immediate\write\@auxout
	{\string\citation{#2}}\fi
\def\@citea{}\@cite{\@for\@citeb:=#2\do
	{\@citea\def\@citea{,}\@ifundefined
	{b@\@citeb}{{\bf ?}\@warning
	{Citation `\@citeb' on page \thepage \space undefined}}
	{\csname b@\@citeb\endcsname}}}{#1}}

\newif\if@cghi
\def\cite{\@cghitrue\@ifnextchar [{\@tempswatrue
	\@citex}{\@tempswafalse\@citex[]}}
\def\citelow{\@cghifalse\@ifnextchar [{\@tempswatrue
	\@citex}{\@tempswafalse\@citex[]}}
\def\@cite#1#2{{$\null^{#1}$\if@tempswa\typeout
	{IJCGA warning: optional citation argument 
	ignored: `#2'} \fi}}

\def\pmb#1{\setbox0=\hbox{#1}
	\kern-.025em\copy0\kern-\wd0
	\kern.05em\copy0\kern-\wd0
	\kern-.025em\raise.0433em\box0}

\def\fnt#1#2{\footnotetext{\kern-.3em
	{$^{\mbox{\scriptsize #1}}$}{#2}}}

\def\fpage#1{\begingroup
\voffset=.3in
\thispagestyle{empty}\begin{table}[b]\centerline{\footnotesize #1}
	\end{table}\endgroup}

\def\runninghead#1#2{\pagestyle{myheadings}
\markboth{{\protect\footnotesize\it{\quad #1}}\hfill}
{\hfill{\protect\footnotesize\it{#2\quad}}}}
\font\tenrm=cmr10
\font\tenit=cmti10 
\font\tenbf=cmbx10
\font\bfit=cmbxti10 at 10pt
\font\ninerm=cmr9

\font\eightrm=cmr8

\def\qed{\hbox{${\vcenter{\vbox{			
   \hrule height 0.4pt\hbox{\vrule width 0.4pt height 6pt
   \kern5pt\vrule width 0.4pt}\hrule height 0.4pt}}}$}}

\begin{document}

\runninghead{Neutrino Physics with Small Extra Dimensions}
{Neutrino Physics with Small Extra Dimensions}

\normalsize\textlineskip
\thispagestyle{empty}
\setcounter{page}{1}

\copyrightheading{}			

\vspace*{0.88truein}

\fpage{1}
\centerline{\bf NEUTRINO PHYSICS WITH SMALL EXTRA DIMENSIONS}
\vspace*{0.37truein}
\centerline{\footnotesize MATTHIAS NEUBERT}
\vspace*{0.015truein}
\centerline{\footnotesize\it Newman Laboratory of Nuclear Studies, 
Cornell University}
\baselineskip=10pt
\centerline{\footnotesize\it Ithaca, New York 14853, U.S.A.}

\vspace*{0.21truein}
\abstracts{
We study neutrino physics in the context of the localized gravity 
model with non-factorizable metric proposed by Randall and Sundrum. 
Identifying the right-handed neutrino with a bulk fermion zero mode, 
which can be localized on the ``hidden'' 3-brane in the 
Randall--Sundrum model, we  obtain naturally small Dirac neutrino 
masses without invoking a see-saw mechanism. Our model predicts
a strong hierarchy of neutrino masses and generically large mixing 
angles.}{}{}


\vspace*{1pt}\textlineskip	

\section{Introduction}

Theories with extra spatial dimensions have received great attention 
recently, when it was shown that they could provide a solution to the 
gauge-hierarchy problem. If space-time is a product of Minkowski 
space with $n$ compact dimensions, with Standard Model fields 
localized in the three extended spatial dimensions (i.e., on a 
3-brane) and gravity propagating in the extra space, then the 
strength of gravity on the 3-brane is governed by an effective Planck 
scale $M_{\rm Pl}^2=M^{n+2}\,V_n$, where $M$ is the fundamental scale 
of gravity and $V_n$ the volume of the compact space.\cite{ADD} If 
this space is sufficiently large, the fundamental scale $M$ can be of 
order 1\,TeV, thus removing the large disparity between the 
gravitational and the electroweak scales. 

An intriguing alternative to the above scenario invokes a 
non-factorizable geometry with a metric that depends on the 
coordinates of the extra dimensions.\cite{RS1} In the simplest 
scenario due to Randall and Sundrum (RS) one considers a single extra 
dimension, taken to be a $S^1/Z_2$ orbifold parameterized by a 
coordinate $y=r_c\phi$, with $r_c$ the radius of the compact 
dimension, $-\pi\le\phi\le\pi$, and the points $(x,\phi)$ and 
$(x,-\phi)$ identified. There are two 3-branes located at 
the orbifold fixed points: a ``visible'' brane at $\phi=\pi$ 
containing the Standard Model fields, and a ``hidden'' brane at 
$\phi=0$. The solution of Einstein's equations for this geometry 
(with carefully tuned brane tensions) leads 
to the non-factorizable metric $ds^2=e^{-2k r_c|\phi|}\,\eta_{\mu\nu}\,
dx^\mu dx^\nu-r_c^2\,d\phi^2$, 
where $x^\mu$ are the coordinates on the four-dimensional surfaces
of constant $\phi$, and the curvature 
parameter $k$ is of order the fundamental Planck scale $M$. In between 
the two branes is a slice of AdS$_5$ space. 

With this setup, the effective Planck scale seen by particles 
confined to four-dimensional space-time is of order the fundamental 
scale $M$. However, the ``warp factor'' $e^{-2k r_c|\phi|}$ in the 
metric has important implications for the masses of particles 
confined to the visible brane. After field renormalization any 
mass parameter $m_0$ in the fundamental theory is promoted into an 
effective mass parameter $m=e^{-k r_c\pi}\,m_0$ governing the physical 
properties of particles on the brane.\cite{RS1} With 
$k r_c\approx 12$ this mechanism produces weak-scale physical masses 
and couplings from fundamental masses and couplings of order the 
Planck scale. As a consequence of the warp factor, the Kaluza--Klein 
excitations of bulk fields have weak-scale mass splittings and 
couplings,\cite{RS2,LyRa,GW1} 
in contrast with the Kaluza--Klein spectra in models with
large extra dimensions.

Resolving the hierarchy problem by introducing extra dimensions poses 
new challenges. In particular, the see-saw mechanism for 
generating small neutrino masses cannot be invoked if the highest 
energy scale governing physics on the visible brane is the weak scale. 
Although several 
four-dimensional alternatives to the see-saw mechanism not requiring 
a high-energy scale have been proposed, 
it would be interesting to find new mechanisms that are intrinsically 
higher dimensional. In the context of models with large extra 
dimensions ideas in this direction have been presented in 
Refs.~\cite{ADDM,ArSc,DS}. They contain a massless Standard Model 
singlet 
propagating in the bulk of the extra compact space, which serves as a 
right-handed neutrino. Then the effective four-dimensional Yukawa 
coupling is suppressed by a volume factor $1/\sqrt{V_n}$, reflecting 
the small overlap between the right-handed neutrino in the bulk and 
the left-handed one on the 3-brane. By construction, this factor 
provides a suppression of neutrino masses of order $v/M_{\rm Pl}$, 
reminiscent of the see-saw mechanism. However, this idea does not work 
in a scenario with small extra dimensions, simply 
because of the lack of a volume suppression factor. 

In a recent work we have investigated the possibility of incorporating 
bulk fermions in the RS model.\cite{us} 
Starting point is the action for a Dirac fermion with mass $m$ 
of order the fundamental scale $M$ propagating in a five-dimensional 
space with the RS metric. We perform the Kaluza--Klein decomposition 
taking into account the orbifold boundary conditions
on the branes, which imply that either 
all left-handed or all right-handed fields must be $Z_2$-odd and thus
vanish on the two 3-branes. Which of these 
choices is realized in nature is a question that cannot be answered 
without understanding in detail the physics on the 3-branes. 
The masses of the Kaluza--Klein fermions are given by 
$m_n=e^{-k r_c\pi} k\,x_n$, 
where $x_n=O(1)$ are roots of some Bessel functions, and the small 
parameter $\epsilon\equiv e^{-k r_c\pi}\sim 10^{-16}$ 
sets the ratio between 
the electroweak and the gravitational scales. Hence the masses $m_n$ 
are of order the weak scale $v$.
Of particular importance to our study are the zero modes supported by
the RS geometry, i.e., solutions with $x_n=0$. The corresponding wave
functions are given by
\begin{equation}\label{zero}
   f_0^{L,R}(\phi) \propto
   \sqrt{\frac{1\pm 2\nu}{1-\epsilon^{1\pm 2\nu}}}\,
   \epsilon^{\mp\nu(|\phi|/\pi-1)} \,,
\end{equation}
where $\nu=m/k$ is a parameter of order unity.
Only one of the zero modes is allowed by the 
orbifold symmetry. This mode exists irrespective of the value of the 
fermion mass $m$ in the five-dimensional theory. Note that for 
$\nu>\frac12$ the right-handed zero mode has a very small wave 
function on the visible brane: $f_0^R(\pi)\propto\epsilon^{\nu-\frac12}$. 
This property allows us to obtain small neutrino masses.

\section{Yukawa interactions and neutrino phenomenology}

We focus first on 
a single fermion generation and consider a scenario where all matter 
and gauge fields charged under the Standard Model gauge group are 
confined to the visible brane at $\phi=\pi$, whereas a gauge-singlet 
fermion field propagates in the bulk. After integration over the 
compact extra dimension we obtain a tower of four-dimensional 
Kaluza--Klein fermions with weak-scale
mass splitting. We choose boundary conditions such that there is 
a right-handed zero mode with wave function $f_0^R(\phi)$ 
given in (\ref{zero}). Only this choice will lead to an interesting
neutrino phenomenology.

We introduce a Yukawa coupling of the bulk fermion with the Higgs 
and lepton fields. With our choice of boundary conditions all 
left-handed Kaluza--Klein modes vanish at the visible brane, so only 
the right-handed modes can couple to the Standard Model fields on the 
brane. Rescaling the Standard Model fields so as to restore 
a canonical normalization, and inserting for the bulk fermion the 
Kaluza--Klein decomposition, we find the action
\[
   S_Y = - \sum_{n\ge 0} \int\!\mbox{d}^4x \left\{ y_n 
   \bar L(x) \widetilde H(x) \psi_n^R(x) + \mbox{h.c.} \right\}
\]
with effective Yukawa couplings $y_n=Y_5\,f_n^R(\pi)$,
where $Y_5$ is naturally of order unity. After electroweak symmetry 
breaking, this Yukawa interaction gives rise to a neutrino mass term 
$\bar\psi_L^\nu\,{\cal M}\,\psi_R^\nu + \mbox{h.c.}$ in the 
basis $\psi_L^\nu=(\nu_L,\psi_1^L,\dots,\psi_n^L)$ and 
$\psi_R^\nu=(\psi_0^R,\psi_1^R,\dots,\psi_n^R)$, with $n\to\infty$.
As a consequence, there will be a mixing of the Standard Model
neutrino $\nu_L$ with the heavy, sterile (with respect to the 
Standard Model gauge interactions) bulk neutrinos $\psi_n^L$. 
In order to obtain a light neutrino we 
need $|y_0|\ll 1$, which requires having a very small wave function 
of the zero mode on the visible brane, i.e., $|f_0^R(\pi)|\ll 1$. But 
this is precisely what happens if the fundamental fermion mass $m$ 
satisfies the condition $m>k/2$.

In order to study the properties of the physical neutrino states we 
diagonalize the squared mass matrix ${\cal M M}^\dagger$. The 
eigenvalues of this matrix are the squares of the physical neutrino 
masses, and the unitary matrix $U$ defined such that 
$U^\dagger {\cal M M}^\dagger U$ is diagonal determines the 
left-handed neutrino mass eigenstates via $\psi_L^\nu
=U\psi_L^{\rm phys}$. We denote by $m_\nu$ the mass of the lightest 
neutrino $\nu_L^{\rm phys}$ and define a mixing angle $\theta_\nu$ 
such that $\nu_L=\cos\theta_\nu\,\nu_L^{\rm phys}+\dots$, where the 
dots represent the admixture of heavy, sterile bulk states. To leading 
order in the small parameter $|y_0|\ll 1$ we obtain (for $\nu>\frac12$)
\[
   m_\nu\sim M \left( \frac{v}{M} \right)^{\nu+\frac12} \,,\qquad
   \tan^2\!\theta_\nu = \frac{1}{2\nu+1}\,\frac{v_0^2 |Y_5|^2}{k^2}
   \,.
\]
The first result is remarkable, as it provides a parametric dependence
of the neutrino mass on the ratio of the electroweak and Planck  
scales that is different from the see-saw relation $m_\nu\sim v^2/M$, 
except for the special case where $\nu=\frac32$. This flexibility
allows us to reproduce a wide range of neutrino masses without any
fine-tuning. For instance, taking $v/M=10^{-16}$, the 
phenomenologically interesting range of $m_\nu$ between $10^{-5}$\,eV 
and 10\,eV can be covered by varying $\nu$ between 1.1 and 1.5. 

The measurement of the invisible width of the $Z^0$ boson, which 
yields $n_\nu=2.985\pm 0.008$ for the apparent number of light 
neutrinos, implies that the mixing angle $\theta_\nu$ 
must be of order a few percent. (Similar constraints follow from 
lepton universality.) For instance, assuming an equal 
admixture of sterile neutrinos for the three generations of light 
neutrinos, we obtain $n_\nu=3\cos^2\!\theta_\nu$ and hence 
$\tan^2\!\theta_\nu=0.005\pm 0.003$. In the context of our model 
this implies that $v_0|Y_5|/k\sim 0.1$, 
which is possible without much fine-tuning. We 
emphasize, however, that it would be unnatural to have the 
dimensionless combination $v_0|Y_5|/k$ much less than unity, so a 
mixing angle $\theta_\nu$ not much smaller than the current 
experimental bound is a generic feature of our scenario, which can be 
tested by future precision measurements.

We now generalize our
mechanism to three neutrino flavors and more than one bulk fermion. 
Interestingly, such a generalization is forced upon us by the 
requirement that the parity anomaly for fermions in an odd number of
dimensions vanish. When an odd number of bulk fermions in five 
dimensions are coupled to a gauge field or gravity, parity invariance
is broken at the quantum level.
To obtain a minimal model that is anomaly free we 
introduce two bulk fermions, 
so there are two massless right-handed zero modes.
In order to explain the atmospheric and solar neutrino anomalies in 
terms of neutrino oscillations one needs two very different 
mass-squared differences: $\Delta m_{21}^2\ll\Delta m_{32}^2$, where 
$\Delta m_{ij}=m_{\nu_i}^2-m_{\nu_j}^2$, and by convention 
$m_{\nu_1}<m_{\nu_2}<m_{\nu_3}$. This requires a minimum of two 
massive neutrinos; however, the third neutrino can be massless. In 
our minimal model this is indeed what happens. 

In order to explore this minimal model in more detail we ignore, for 
simplicity, the heavy Kaluza--Klein excitations of the bulk fermions
and focus only on the zero modes. The admixture 
of weak-scale sterile neutrino states must be strongly suppressed. It
is natural to allow for the possibility that the two bulk fermions 
have different masses $m_1>m_2$ (of order the Planck scale) in the 
fundamental theory, and that they couple with similar strength to the 
three left-handed neutrino flavors. Diagonalizing the resulting 
neutrino mass matrix
$\bar\psi_L^\nu\,{\cal M}\,\psi_R^\nu+\mbox{h.c.}$ in the truncated 
basis $\psi_L^\nu=(\nu_e^L,\nu_\mu^L,\nu_\tau^L)$ and 
$\psi_R^\nu=(\psi_0^{R,1},\psi_0^{R,2})$ to leading order in 
$\epsilon$ we find that the physical neutrino mass eigenstates 
comprise a massless left-handed neutrino $\nu_1$, a very light Dirac 
neutrino with mass $m_{\nu_2}$, and a light Dirac neutrino with mass
$m_{\nu_3}$, where
\[
   m_{\nu_2}^2 \sim M^2 \left( \frac{v}{M} \right)^{2\nu_1+1} \,,
   \qquad
   m_{\nu_3}^2 \sim M^2 \left( \frac{v}{M} \right)^{2\nu_2+1} \,.
\]
An interpretation of the solar neutrino anomaly in terms of neutrino 
oscillations based on the MSW effect yields values of 
$\Delta m_{21}^2$ in the range $10^{-6}$--$10^{-5}$\,eV$^2$, whereas 
oscillations in vacuum would require a smaller value of order 
$10^{-10}$\,eV$^2$. Such masses can be reproduced in our 
model by setting $\nu_1\approx 1.34$--1.37 and $\nu_1\approx 1.5$, 
respectively. An explanation of the atmospheric neutrino anomaly in 
terms of neutrino oscillations yields $\Delta m_{32}^2$ in the range 
$5\cdot 10^{-4}$--$6\cdot 10^{-3}$\,eV$^2$, which we can 
reproduce by taking $\nu_2\approx 1.27$--1.29. In other words, we 
can understand the observed hierarchy of the experimentally favored 
neutrino masses in terms of a small difference of the bulk fermion
masses in the fundamental theory.

Despite the fact that a strong neutrino mass hierarchy is a generic
feature of our model, the mixing matrix $U$ relating the neutrino 
flavor and mass eigenstates does not contain any small parameter. 
Defining $\nu_f=\sum_{i=1}^3 U_{fi}\,\nu_i$ we find that all the 
entries $U_{fi}$ are of order unity. A mixing matrix of this type,
which lacks the strong hierarchy of the quark mixing matrix, is 
compatible with the experimental constraints on the neutrino mixing 
angles.

\section{Conclusions}

We have presented an intriguing scenario for the origin of neutrino
masses and mixings in the context of the Randall--Sundrum 
extra-dimension model with non-factorizable geometry. Bulk fermion 
states in this model have chiral zero modes localized on one of the 
two 3-branes, depending on the choice of orbifold boundary conditions.
If the localization is done on the ``hidden'' brane, then these states
have a wave function on the 
visible brane that is power-suppressed in the ratio of the weak scale 
to the fundamental Planck scale. Coupling 
the Higgs and left-handed lepton fields of the Standard Model, 
localized on the visible brane, with a bulk right-handed neutrino 
provides a new mechanism for obtaining small neutrino masses. 
Remarkably, this mechanism leads to a generalization of the see-saw 
formula with a different parametric dependence on the ratio $v/M$, 
which can easily reproduce neutrino masses in the range $10^{-5}$\,eV 
to 10\,eV. With an even number of bulk fermions 
one can obtain viable models of neutrino flavor oscillations, which 
naturally predict a mass hierarchy and a neutrino mixing matrix not 
containing any small parameter.

\nonumsection{Acknowledgements}
\noindent
It is a pleasure to thank Yuval Grossman for collaboration on the 
subject of this talk. This work was supported in part by the National 
Science Foundation.

\nonumsection{References}

\end{document}